\documentclass[prb,twocolumn,floatfix,notitlepage,superscriptaddress,longtable]{revtex4-2}
\usepackage{amsmath}
\usepackage{lipsum} 
\usepackage{verbatim}
\usepackage{epsfig}
\usepackage{subfigure}
\usepackage{graphicx}
\usepackage{amsfonts}
\usepackage[figuresright]{rotating}
\usepackage{amssymb}
\usepackage{amsmath}
\usepackage[normalem]{ulem}
\usepackage{psfrag}
\usepackage{esint} 
\usepackage{bm}
\usepackage[colorlinks,linkcolor=blue,anchorcolor=blue,citecolor=blue,urlcolor=blue]{hyperref}
\usepackage[version=4]{mhchem}
\usepackage[svgnames]{xcolor}

\usepackage{graphicx} 
\usepackage{xcolor}   
\usepackage{booktabs}
\usepackage{tikz}
\usetikzlibrary{positioning, shapes.geometric, arrows.meta, calc, fit, shadows}
\usetikzlibrary{arrows.meta} 
\usepackage{pgfplots}
\pgfplotsset{compat=1.18}    

\def\be{\begin{equation}} \def\ee{\end{equation}}
\def\bea{\begin{eqnarray}} \def\eea{\end{eqnarray}}

\renewcommand{\vec}[1]{\mathbf{#1}}

\def\bpm{\begin{pmatrix}} \def\epm{\end{pmatrix}}

\renewcommand{\d}{\text{d}}

\newcommand{\nin}{n_{\text{in}}}

\DeclareMathOperator{\Tr}{Tr}

\definecolor{Qicolor}{RGB}{3, 136, 252}

\makeatletter
\newcommand*{\balancecolsandclearpage}{%
  \close@column@grid
  \clearpage
}
\makeatother

\begin{document}
\title{Attention in Krylov Space: Transformer-Based Extrapolation of Lanczos Coefficients}
\author{Zihao Qi}
\email{zq73@cornell.edu}
\affiliation{Department of Physics, Cornell University, Ithaca, NY 14853, USA.}

\author{Christopher Earls}
\affiliation{Center for Applied Mathematics, Cornell University, Ithaca, NY 14853, USA.}

\begin{abstract}
The Universal Operator Growth Hypothesis formulates time evolution of operators through Lanczos coefficients. In practice, however, numerical instability and memory cost limit the number of coefficients that can be exactly computed. In response to these challenges, the standard approach relies on fitting early coefficients to asymptotic forms, but such procedures can miss subleading, history-dependent structures in the coefficients that subsequently affect reconstructed observables. In this work, we treat the Lanczos coefficients as a causal time sequence and introduce a transformer-based model to autoregressively predict future Lanczos coefficients from short prefixes. For classical and quantum chaotic systems, our model outperforms asymptotic fits in both coefficient extrapolation and physical observable reconstruction, and achieves an order-of-magnitude reduction in error. The model also accurately extrapolates coefficients in integrable regimes, where no universal asymptotic fit exists. Remarkably, our model transfers across system sizes: it can be trained on smaller systems and then be used to extrapolate coefficients on a larger system \emph{without retraining}. By probing the learned attention patterns and performing targeted attention ablations, we identify portions of the coefficient history that are most influential for accurate forecasts. Our results demonstrate that modern sequence models can serve as practical surrogates for probing operator dynamics deep in Krylov space, where brute-force Lanczos iteration can be computationally prohibitive.
\end{abstract}

\maketitle

\section{Introduction}
Thermalization is a defining feature in generic interacting quantum many-body systems. While time evolution of the full system is unitary, local observables typically thermalize, i.e. relax towards values consistent with a thermal equilibrium state, rendering microscopic details of the initial state inaccessible to local probes~\cite{Srednicki1994, Deutsch1991, Rigol2008,DAlessio2016,GogolinEisert2016,EisertFriesdorfGogolin2015, Polkovnikov2011RMP, Deutsch2018RPP}. This emergent, irreversible loss of information is due to the rapid delocalization, or ``scrambling'', of quantum information across many degrees of freedom, and it is central to the studies of quantum many-body dynamics and quantum information science~\cite{Maldacena2016,HaydenPreskill2007,SekinoSusskind2008,Hosur2016,Swingle2018, UOGH}. 

One convenient approach to studying information spreading is to work directly in the Heisenberg picture, treating time evolution as operator dynamics in an operator Hilbert space. Recently, Parker et al.~\cite{UOGH} proposed a framework to quantify the growth of operator complexity in both classical and quantum systems. In this formalism, Heisenberg time evolution of a local operator can be mapped to a hopping problem on a semi-infinite Krylov chain, where positions far away from the origin correspond to more spatially delocalized and therefore more complex operators~\cite{UOGH, Caputa2022GeometryKrylov, Nandy2025PhysRepKrylovReview}. The formalism has proven to be remarkably successful for studying dynamics in a wide range of physical systems, providing insights into classical chaos~\cite{UOGH,Barbon2019}, open quantum systems~\cite{Liu2023, operator_growth_open, operator_growth_open2, operator_growth_open3, operator_growth_open4, pratik1, pratik2, solvable_model_lindbladian}, and various quantum spin models~\cite{UOGH, Yates_2020, DHS, xiangyu_statmechanism_spin, spin_model, nonhermitian_spinchain, estimate_timescale_spin, pseudomode_extrapolate_spin, approx_Greens_extrap_and_cutoff_spin, MBL, Yates2_staggering, staggering2, Yates_Floquet_staggering}. Krylov complexity has also attracted considerable recent interest due to its connection with holography~\cite{UOGH, Caputa2022GeometryKrylov, Barbon2019, Rabinovici2021KrylovEdge, holography1, holography2, qft_staggering, holography3}.

Central to operator growth dynamics are the Lanczos coefficients $\{b_n\}$, which serve as hopping amplitudes along the Krylov chain and result from iterative applications of the Liouvillian to an initial operator~\cite{UOGH}. The asymptotic behavior of Lanczos coefficients encodes important information about the underlying dynamics: sub-linear growth such as $b_n \sim \sqrt{n}$ indicates integrability or constrained dynamics~\cite{UOGH, operator_growth_open2, integrability}, while chaotic systems exhibit a linear scaling $b_n \sim \alpha n + \gamma$, with a logarithmic correction in 1D~\cite{UOGH,Heveling2022ProbeUOGH}. The growth rate $\alpha$, in particular, constrains the rate of information scrambling and provides bounds related to Lyapunov growth and out-of-time-order correlators (OTOCs)~\cite{UOGH}.

While the Lanczos formalism provides a compact representation of operator dynamics, its practical implementation can be challenging.
First, computing long sequences of Lanczos coefficients is notoriously susceptible to numerical instabilities arising from loss of orthogonality due to finite-precision arithmetic~\cite{loss_orthogonality}.
Although there exist techniques that can circumvent such instability~\cite{edge_krylov_space, reorthogonalization, partial_orthogonalization, ParlettScott1979SelectiveOrthog, dynamics_oneandtwo_d, lychkovskiy}, generating long sequences of Lanczos coefficients still requires repeated applications of the Liouvillian. In quantum many-body systems, the Hilbert space dimension scales exponentially with the system size, causing the memory and runtime costs of the Lanczos algorithm to rapidly grow prohibitive for larger systems.

\begin{figure}[t]
    \centering
    \resizebox{0.8\linewidth}{!}{%
        \begin{tikzpicture}
\begin{axis}[
    width=10cm,
    height=8cm,
    title={},
    xlabel={$n$},
    ylabel={$b_n$},
    xlabel style={font=\huge}, 
    ylabel style={font=\huge},
    xtick=\empty,
    ytick=\empty,
    xmin=0, xmax=22,
    ymin=0, ymax=12,
    legend pos=north west,
    legend style={draw=none, fill=none, nodes={scale=0.9}},
    axis line style={thick}
]

\addplot[
    domain=1:21, 
    samples=2, 
    dashed, 
    gray, 
    thick
] {0.5*x + 1};
\addlegendentry{\large Linear Fit ($b_n \sim \alpha n$)}

\addplot[
    only marks,
    mark=*,
    blue,
    mark size=2.5pt,
    error bars/.cd,
    y dir=minus, 
    y explicit,
    error bar style={red, thick}
] 
table[x=x, y=y, y error expr=\thisrow{y} - (0.5*\thisrow{x} + 1)] {
    x   y
    1   1.2
    2   2.5
    3   2.0
    4   3.8
    5   4.5
    6   3.2
    7   5.4
    8   5.5
    9   6.2
    10  4.5
    11  5.8
    12  7.5
    13  8.2
    14  7.0
    15  10.0
    16  9.8
    17  10.5
    18  9.0
    19  10.3
    20  11.5
    21  10.8
};

\node[anchor=west, font = \large] (actual) at (axis cs: 12, 3) {Actual Coefficients};
\draw[-{Stealth}, thick] (actual.west) -- (axis cs: 10, 4.4);

\node[anchor=west, align=left, font = \large] (dev) at (axis cs: 16.5, 7.3) {Deviations};
\draw[-{Stealth}, thick] (dev.west) -- (axis cs: 14, 7.3);

\end{axis}
\end{tikzpicture}%
    }
    \caption{Schematic illustration of a typical sequence of Lanczos coefficients. The dashed line indicates an asymptotic linear fit, while the red segments depict residual deviations that are not captured by simple asymptotic fits. The deviations are systematic subleading structures and can strongly affect reconstructed dynamics~\cite{approx_Greens_extrap_and_cutoff_spin, Yates_2020, Yates2_staggering, staggering2, Yates_Floquet_staggering, subleading, subleading2, pseudomode_extrapolate_spin, dynamics_oneandtwo_d, staggering_cft}. }
    \label{fig:deviation}
\end{figure}
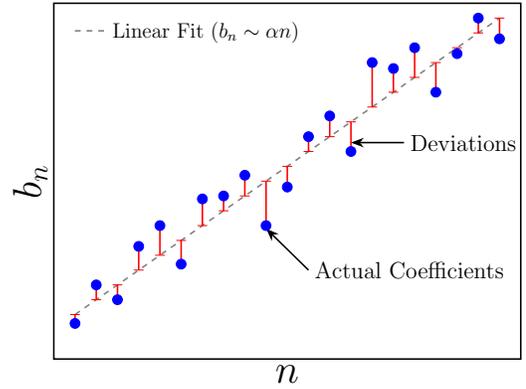

Due to these practical constraints, one can often only reliably compute a very limited number of Lanczos coefficients in typical quantum many-body systems. Standard practice therefore relies on \textit{extrapolating} future coefficients, via a fit of the asymptotic form to coefficients computed in an initial window~\cite{MBL, pseudomode_extrapolate_spin, approx_Greens_extrap_and_cutoff_spin, Yates_2020, dynamics_oneandtwo_d, Heveling2022ProbeUOGH, diffusion_constant, Yates2_staggering, Yates_Floquet_staggering}. 
However, ground-truth coefficients often exhibit systematic deviations from the  asymptotic trend, such as subleading corrections, oscillations, and model-dependent structures, as illustrated in Fig.~\ref{fig:deviation}. 
The resulting deviations should not be viewed as noise; rather, such deviations can have concrete and important physical consequences. For example, the even-odd staggering around the linear asymptote encodes crucial information regarding long-lived edge modes of the system~\cite{Yates_2020, Yates2_staggering, staggering2, Yates_Floquet_staggering}. The non-linear, sub-leading terms have also been shown to be directly related to low-frequency behaviors of the spectral functions~\cite{subleading, subleading2, subleading3}. Simple linear extrapolation is unable to respect these intricate patterns, leading to errors in reconstructed autocorrelation functions and other physical quantities.

The numerical bottlenecks of exact Lanczos iterations, together with the limitations of linear fits, motivate a complementary, data-driven approach: given only a short prefix of coefficients $\{b_1, b_2, \dots, b_{\nin} \}$, can we reliably predict Lanczos coefficients beyond the range accessible to direct computation, using a learned surrogate model? Similar short-window to long-horizon forecasting based on machine-learning models has recently proven effective in other contexts, such as predicting long-time dissipative trajectories from short-time evolution~\cite{RodriguezKananenka2021CNN,RodriguezKananenka2024TransformerDissipative} and forecasting dynamics in strongly entangled many-body systems~\cite{Kaneko2025DMD}. More broadly, machine learning approaches have emerged as powerful complements to traditional numerical techniques across quantum many-body physics~\cite{NNQS, NNQS1, NNQS3, FNO, Carleo2017, Carrasquilla2017, transformer_gauge_lattice_theory, transformer_quantum_state, transformer_quantum_state2, transformer_tomography,Carleo2019MLPhysSciences, Carrasquilla2020MLQuantumMatter, attention_manybody, attention_manybody2, attention_manybody3}.

In this work, we formulate the extrapolation of Lanczos coefficients as being a causal sequence-prediction problem and adopt a transformer-based architecture~\cite{Vaswani2017}. Transformers have been remarkably successful across a broad range of sequence modeling tasks, including natural language processing~\cite{Vaswani2017, Devlin2018} and time-series forecasting~\cite{Zhou2021}. By utilizing a ``self-attention'' mechanism, the model assigns a measure of relative importance to \textit{every} coefficient that precedes and contributes to the prediction of the next one. This self-attention architecture allows the transformer model to capture subtle, history-dependent structures and long-range correlations in Lanczos coefficients that standard asymptotic fit extrapolations unavoidably miss.

We benchmark our approach on three paradigmatic models of classical chaos, quantum chaos, and integrability: the XYZ spin top~\cite{LiuMuellerXYZ}, the transverse-field Ising model with integrability-breaking field~\cite{TFIM}, and the Heisenberg XXZ Model~\cite{XXZ}. For chaotic systems, we demonstrate that our proposed transformer model is able to forecast Lanczos coefficients and reconstruct physical quantities more accurately than baseline fits based on asymptotic forms.
Our proposed model can be transferred across system sizes \textit{without retraining}: trained on Lanczos sequences in smaller systems, which are cheap to obtain, the model accurately extrapolates coefficients for larger system sizes. Our approach opens up the possibility of using modern sequence prediction models as practical surrogates for probing dynamics deep within Krylov spaces, where brute-force Lanczos iteration is computationally prohibitive; thereby enabling more reliable access to long-time operator dynamics from limited input.

The remainder of this work is organized as follows. In Sec.~\ref{sec:lanczos}, we briefly review the formalism of operator growth in terms of Lanczos coefficients and corresponding Krylov basis. Sec.~\ref{sec:learning} details the training data, learning objective, as well as our model architecture. In Sec.~\ref{sec:results}, we demonstrate the proposed transformer's superiority in Lanczos coefficient forecasting and observable reconstruction, as compared with state-of-the-art linear asymptotic predictions. We analyze the learned attention structure and conduct an ablation study in Sec.~\ref{sec:discussion}, and summarize the work in Sec.~\ref{sec:conclusion}.

\section{Review of Lanczos Formalism for Operator Growth \label{sec:lanczos}}

In generic non-integrable many-body systems, initially simple operators typically grow more complex under Heisenberg time evolution.
The Universal Operator Growth Hypothesis (UOGH)~\cite{UOGH} relates this growth to the large-$n$ behavior of Lanczos coefficients generated from the Liouvillian. In this section, we follow Ref.~\cite{UOGH} and briefly review the formalism for quantum systems. The derivations carry over to classical systems almost verbatim~\cite{UOGH}; we discuss the differences in Appendix~\ref{app:classical_lanczos}.

Consider a quantum system described by a Hamiltonian $H$. We denote operators as states in an operator Hilbert space, using the notation $\hat{A} = |A)$. Throughout this work, we endow the operator Hilbert space with the (infinite-temperature) Hilbert-Schmidt inner product, 
\begin{equation}
    (A|B) = \Tr(\rho_\infty A^\dagger B) =  \frac{1}{\Tr(\mathbb{I})} \Tr(A^\dagger B),
\end{equation}
where $\mathbb{I}$ is the identity matrix. In the Heisenberg picture, operators evolve in time according to the Heisenberg equation of motion:
\begin{equation}
    -i \partial_t |O) = [H, O] = \mathcal{L} |O),
\end{equation}
where we have defined the \textit{Liouvillian} \textit{super}operator as $\mathcal{L}(\cdot) = [H, \cdot]$. The time-evolved operator is thus:
\begin{equation}
  |O(t)) =   e^{i\mathcal{L}t} |O_0) = \sum_n \frac{(it)^n}{n!} \mathcal{L}^n | O_0),
\end{equation}
and it lies in the Krylov subspace $\mathcal{K}$, spanned by repeated actions of $\mathcal{L}$ on the initial operator, $\mathcal{K} = \text{span} \{ \mathcal{L}^n |O_0)\}_{n \geq 0}$.

The Krylov subspace, from this definition, is not yet orthonormal.
To analyze the dynamics within this subspace, we apply the Lanczos algorithm to produce an orthonormal basis of operators. This procedure generates a sequence of positive numbers $\{b_n \}$ (the Lanczos coefficients) and a sequence of orthonormal operators $\{|O_n)\}$ (the Krylov basis).

More concretely, assume that the initial operator $O_0$ is properly normalized, $(O_0 | O_0)=1$. We define $|A_1) := \mathcal{L} |O_0) = [H, O_0]$ and $b_1 := \sqrt{\left( A_1 | A_1 \right)}$.
We then normalize $A_1$ to find the next basis operator $|O_1) := b_1^{-1} |A_1)$. The subsequent coefficients and basis elements for $n \geq 2$ are defined iteratively as:
\begin{align}
    |A_n) &= \mathcal{L} |O_{n-1}) - b_{n-1} |O_{n-2}) \nonumber \\
    b_n &= \sqrt{\left( A_n | A_n \right)} \nonumber \\
    |O_n) &= b_n^{-1} |A_n).
    \label{eq:lanczos}
\end{align}

One can check by induction that the resulting Krylov basis  $\{ |O_n) \}$ is orthonormal: $\left( O_n | O_m \right) = \delta_{nm}$.  In this basis, the Liouvillian is tri-diagonal, with diagonal entries being zero, and subdiagonal entries being the Lanczos coefficients:

\begin{equation}
   \mathcal{L} = \begin{pmatrix}
         0 & b_1 & 0 & 0  \\
         b_1 & 0 & b_2 & 0 \\
         0 & b_2 & 0 & b_3 \\
         0 & 0 & b_3 & 0 &  \\
          & & & & \ddots
    \end{pmatrix}.
    \label{eq:L_tridiag}
\end{equation}

The time-evolved operator $O(t)$ can now be decomposed in the Krylov basis, with time-dependent coefficients:
\begin{equation}
    |O(t)) = \sum_n \phi_n(t) |O_n).
    \label{eq:Otexpansion}
\end{equation}
Importantly, under this decomposition, the Heisenberg equation can be mapped to a single-particle hopping problem on a 1D semi-infinite chain (the Krylov chain), governed by the following equation of motion:
\begin{equation}
    -i \partial_t \phi_n = b_n \phi_{n-1} + b_{n+1} \phi_{n+1},
    \label{eq:hoppingeq}
\end{equation}
with initial conditions $\phi_n(t=0) = \delta_{n0}$ and $\phi_{-1}(t) = 0$. The Lanczos coefficients $b_n$ therefore act as nearest-neighbor hopping amplitudes on the semi-infinite chain. The average position of the wavefunction along the Krylov chain,
\begin{equation}
    K(t) = \sum_n n |\phi_n(t)|^2,
    \label{eq:krylov_complexity}
\end{equation}
is called the Krylov complexity. Intuitively, $K(t)$ encodes the spatial extent of the wavefunction, and it provides a quantitative measure of the growth of complexity for the initial operator $|O_0)$ under time evolution.

The scaling of $b_n$ with $n$ encodes important information about the nature of operator growth in the system. Specifically, the Universal Operator Growth Hypothesis (UOGH) posits that under generic \textit{chaotic} dynamics, the Lanczos coefficients $b_n$ follow specific asymptotic forms:
\begin{align}
    b_n \sim \alpha n/\log(n) + \gamma \,\,\, \,\,\,d=1; \nonumber \\ 
    b_n \sim \alpha n + \gamma \,\,\,\,\,\, d\neq 1,
\label{eq:UOGH}
\end{align}
where $d$ is the spatial dimension of the system. The growth rate $\alpha$ is directly related to the spreading of information in the system~\cite{UOGH}.

In general, both the Lanczos coefficients $b_n$ and the Krylov basis $|O_n)$ are required to fully construct the time-evolved operator. However, the Lanczos coefficients by themselves are also directly related to physical observables, such as the auto-correlation function of the operator $|O_0)$, defined as:
\begin{equation} 
C(t) = (O_0 | O(t)).
\end{equation} 

Using the expansion of $O(t)$, Eq.~\ref{eq:Otexpansion}, we find that the autocorrelation function can also be written as:
\begin{equation} 
C(t) = \sum_n \phi_n(t) (O_0 | O_n) = \phi_0(t),
\label{eq:autocorrelation}
\end{equation}
corresponding to the amplitude of the particle remaining at the first site of the Krylov chain. 

In many settings, the physically relevant quantities (such as the autocorrelation $C(t)$ and Krylov complexity $K(t)$) require knowledge of $\{b_n\}$
out to moderately large values of $n$. At the same time, direct computation of $\{b_n\}$ becomes increasingly costly
with $n$ and is often strongly limited by memory constraints and numerical instability. Asymptotic fits based on a limited number of early coefficients, on the other hand, necessarily fail to capture the physically significant subleading structures. Due to these limitations, we switch to a data-driven framework. In the following section, we detail our machine learning approach that treats the Lanczos sequence as a structured time series and performs causal, autoregressive extrapolation directly in Krylov space.

\section{Data Representation and Model Architecture \label{sec:learning}}
We first specify the dataset, learning target, as well as model architecture used in this work. To generate training data, we fix an initial operator $O_0$ and sample Hamiltonians $H^{(j)}$ from a chosen model family (details in Sec.~\ref{sec:results}) independently and identically. For each sampled $H^{(j)}$, we apply the Lanczos procedure (Eq.~\ref{eq:lanczos}) to generate a sequence of Lanczos coefficients $\{b_n^{(j)}\}$. Full reorthogonalization~\cite{edge_krylov_space} is applied at each step to ensure numerical stability. Our training dataset therefore consists of $N$ sequences of Lanczos coefficients 
\begin{equation}
   \vec{b}^{(j)} = \{b_1^{(j)}, b_2^{(j)} ..., b_{T}^{(j)} \}, \, \, j \in \{1, 2, ..., N\},
\end{equation}
where the sequence labeled by the index $j$ is generated by dynamics under $H^{(j)}$. Here $T$ denotes the total length of each sequence of Lanczos coefficients.

Next we process the training data, as modeling $\{b_n\}$ directly can be numerically inconvenient due to the large dynamical range. We therefore consider an equivalent formulation in terms of \textit{differences} between neighboring coefficients, defined as
\begin{equation}
    \Delta b_n := b_n - b_{n-1}.
\end{equation}
Here we adopt the convention $b_0 \equiv 0$, so that $\Delta b_1 = b_1$.

Using $\Delta b_n$ instead of $b_n$ yields a scale-stabilized target and therefore allows for more stable training and extrapolation. Under UOGH, the raw coefficients grow without bound, whereas $\Delta b_n$ approaches an $n$-independent constant in the asymptotic limit. The original coefficients can be reconstructed via a cumulative sum,
\begin{equation}
    b_n = \sum_{j=1}^{n} \Delta b_j,
    \label{eq:cum_sum}
\end{equation}

With the data prepared and processed, we now formulate our problem as an autoregressive next-step prediction task. Specifically, given the first $\nin$ Lanczos coefficients (or equivalently, $\{\Delta b_1, \Delta b_2, ..., \Delta b_{\nin} \}$), the goal is to train a machine-learning model to predict coefficient differences $\Delta b_{\nin +1}, \Delta b_{\nin +2}, \dots $, sequentially, and then reconstruct future $b_n$'s via Eq.~\ref{eq:cum_sum}. To this end, we adopt a transformer architecture. Transformers excel at capturing long-range dependencies and structured correlations in sequential data, through a learned self-attention mechanism. The transformer architecture has proven useful in a wide range of physical problems, such as representing many-body wavefunctions~\cite{transformer_quantum_state, transformer_quantum_state2, transformer_wavefunction}, sampling lattice gauge theories~\cite{transformer_gauge_lattice_theory}, and reconstructing quantum states~\cite{transformer_tomography}.

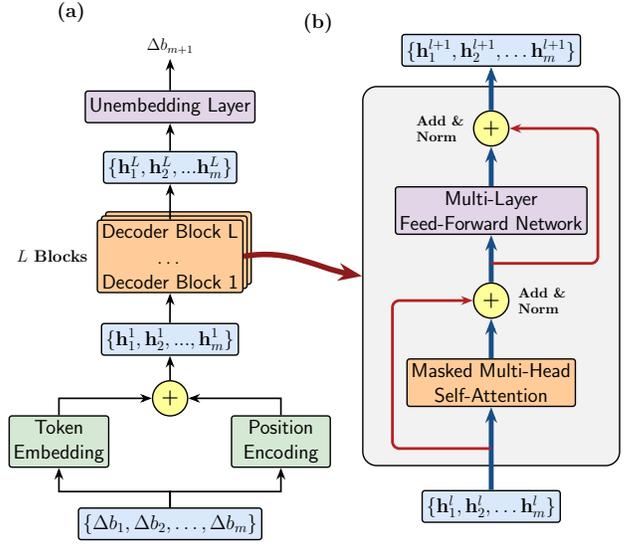
\begin{figure}[t]
    \centering
    \resizebox{\linewidth}{!}{%
\definecolor{boxblue}{RGB}{218, 232, 252}
\definecolor{boxgreen}{RGB}{213, 232, 212}
\definecolor{boxorange}{RGB}{255, 204, 153}
\definecolor{boxpurple}{RGB}{225, 213, 231}
\definecolor{addyellow}{RGB}{255, 255, 153}
\definecolor{arrowblue}{RGB}{22, 79, 134}
\definecolor{arrowred}{RGB}{178, 34, 34}
\definecolor{panelgray}{RGB}{242, 242, 242}


\begin{tikzpicture}[
    font=\sffamily\Large,
    line width=1pt,
    base/.style={draw, rounded corners=2pt, align=center, minimum height=0.8cm, minimum width=2.5cm},
    blockstack/.style={base, fill=boxorange, double copy shadow={shadow xshift=3pt, shadow yshift=3pt, fill=boxorange}},
    add/.style={draw, circle, fill=addyellow, minimum size=0.6cm, font=\Large\bfseries},
    stdarrow/.style={-{Stealth[length=3mm]}, line width=1.2pt},
    thickblue/.style={-{Stealth[length=4mm]}, color=arrowblue, line width=3.5pt},
    thickred/.style={-{Stealth[length=3mm]}, color=arrowred, line width=2pt}
]


\node[base, fill=boxblue] (in) {$\{\Delta b_1, \Delta b_2, \dots, \Delta b_m \}$};
\node[base, fill=boxgreen, above left=1.0cm and -0.8cm of in] (token) {Token\\Embedding};
\node[base, fill=boxgreen, above right=1.0cm and -0.8cm of in] (pos) {Position\\Encoding};

\node[add, above=2.3cm of in] (add_emb) {+};
\node[base, fill=boxblue, above right=0.75cm and -2.08cm of add_emb] (h0) {$\{\mathbf{h}^1_1, \mathbf{h}^1_2, ..., \mathbf{h}_m^1 \}$};

\node[blockstack, above=2.2cm of add_emb, minimum width=3.5cm,] (dec) {Decoder Block L\\ \dots \\ Decoder Block 1};
\node[left=0.1cm of dec, align=right, font=\large\bfseries] {$L$ Blocks};
\node[base, fill=boxblue, above=0.9cm of dec] (hL) {$\{\mathbf{h}^L_1, \mathbf{h}^L_2, ... \mathbf{h}^L_m \}$};

\node[base, fill=boxpurple, above=0.7cm of hL, minimum width=3.5cm] (linear) {Unembedding Layer};
\node[above=0.8cm of linear, align=center, font=\large] (prob) {$\Delta b_{m+1}$};

\draw[stdarrow] (in.north) -- ++(0,0.4) -| (token.south);
\draw[stdarrow] (in.north) -- ++(0,0.4) -| (pos.south);
\draw[stdarrow] (token.north) |- (add_emb.west);
\draw[stdarrow] (pos.north) |- (add_emb.east);
\draw[stdarrow] (add_emb.north) -- (h0.south);
\draw[stdarrow] (h0.north) -- (dec.south);
\draw[stdarrow] (dec.north) -- (hL.south);
\draw[stdarrow] (hL.north) -- (linear.south);
\draw[stdarrow] (linear.north) -- (prob.south);

\node[above left=0.16cm and 1.3 cm of prob, font=\bfseries, font=\Large\bfseries] {(a)};


\node[draw, fill=panelgray, rounded corners=10pt, minimum width=6.5cm, minimum height=9.6cm,  below right=0.5cm and 3.0cm of dec.east, anchor=west] (bg) {};

\node[base, fill=boxorange, above=1.5cm of bg.south, minimum width=4cm] (mha) {Masked Multi-Head\\Self-Attention};
\node[add, above=1.0cm of mha] (add1) {+};
\node[right=0.1cm of add1, align=left, font=\normalsize \bfseries] {Add \&\\Norm};

\node[base, fill=boxpurple, above=1.2cm of add1, minimum width=4cm] (mlp) {Multi-Layer\\ Feed-Forward Network};
\node[add, above=1.0cm of mlp] (add2) {+};
\node[left=0.1cm of add2, align=left, font=\normalsize \bfseries] {Add \&\\Norm};

\node[base, fill=boxblue, below=0.5cm of mha, yshift=-1.6cm] (bin) {$\{\mathbf{h}_1^l, \mathbf{h}_2^l,\dots \mathbf{h}_m^l \}$};
\node[base, fill=boxblue, above=0.3cm of add2, yshift=0.8cm] (bout) {$\{\mathbf{h}_1^{l+1}, \mathbf{h}_2^{l+1},\dots \mathbf{h}_m^{l+1} \}$};

\draw[thickblue] (bin.north) -- (mha.south);
\draw[thickblue] (mha.north) -- (add1.south);
\draw[thickblue] (add1.north) -- (mlp.south);
\draw[thickblue] (mlp.north) -- (add2.south);
\draw[thickblue] (add2.north) -- (bout.south);

\draw[thickred, rounded corners=5pt] ($(bin.north)!0.5!(mha.south)$) -- ++(-2.5,0) |- (add1.west);

\draw[thickred, rounded corners=5pt] ($(add1.north)!0.4!(mlp.south)$) -- ++(2.7,0) |- (add2.east);

\node[above left=-0.1cm and 1.6cm of bout, font=\Large\bfseries] {(b)};


\draw[line width=4pt, color=arrowred!80!black, -{Stealth[length=6mm]}] (dec.east) to[out=10, in=170] (bg.west);

\end{tikzpicture}
}%
    \caption{\textbf{(a)} Illustration of the Decoder-only transformer architecture. The scalar inputs $\{ \Delta b_1 \dots\Delta b_{m} \}$ are mapped to $d_\text{model}$-dimensional token embeddings and augmented by positional encodings to become hidden states $\{\mathbf{h}_1^1 \dots \mathbf{h}_m^1\}$, which pass through $L$ decoder blocks. The final hidden state is projected to the next token prediction $\Delta b_{m+1}$ by the unembedding (output) layer.
    \textbf{(b)} Internal structure of the $l$-th decoder block, showing the masked multi-head self-attention mechanism, residual connections, and the feed-forward network.}
    \label{fig:transformer}
\end{figure}

The architecture of our transformer model is illustrated in Fig.~\ref{fig:transformer}(a). Each scalar input $\Delta b_n \in \mathbb{R}$, where $n \in (1, 2, \dots, m)$ and $m$ is the total length of input window, is first mapped to a $d_{\text{model}}$-dimensional embedding vector via a learned, affine token embedding,
\begin{equation}
\mathbf{e}_n = W_E \Delta b_n + \mathbf{b_E},\quad
W_E \in \mathbb{R}^{d_{\text{model}} \times 1},  \mathbf{b_E} \in \mathbb{R}^{d_{\text{model}}}.
\end{equation}

Note that the embedding operation is independent of token index $n$. The bold-font $\vec{b}_E$ denotes the bias vector, and is not to be confused with Lanczos coefficients, which are scalars. In addition to the embedding, we add a positional encoding $\mathbf{p}_n \in \mathbb{R}^{d_{\text{model}}}$ to inject information about the coefficient index $n$~\cite{Vaswani2017}. This allows the model to distinguish early coefficients from later ones; and this is essential for sequence modeling. The resulting embedded vector $\mathbf{h}_n^1$, defined as
\begin{equation}
\mathbf{h}_n^1 = \mathbf{e}_n + \mathbf{p}_n,
\end{equation}
serves as the initial embedding-space representation for $\Delta b_n$.

Next, the embedded input vectors, $\{\vec{h}_1^1, \vec{h}_2^1 \dots \vec{h}_{m}^1 \}$, are sequentially processed through $L$ decoder blocks. The subscript and superscript denote indices for the token and decoder block, respectively. The internal workings of the $l$-th decoder block is shown in Fig.~\ref{fig:transformer}(b). 

Within each decoder block, the vectors first pass through a masked self-attention mechanism with $H$ attention heads. In each attention head, labeled by $i \in \{1, \dots,H\}$, the input vectors $\vec{h}_n^l$ are first multiplied by three learnable matrices $W_{K, i}^l, W_{Q,i}^l,$ and $W_{V, i}^l$:
\begin{equation}
 \vec{k}_{n,i}^l = W_{K,i}^l \vec{h}_n^l; \quad \vec{q}_{n,i}^l = W_{Q,i}^l \vec{h}_n^l; \quad \vec{v}_{n,i}^l = W_{V,i}^l \vec{h}_n^l,
\end{equation}
where $W_{K, i}^l, W_{Q,i}^l, W_{V, i}^l \in \mathbb{R}^{d_k \times d_\text{model}}$, and $d_k := d_\text{model}/H$ is the reduced dimension in each head. The resulting vectors $\vec{k}_{n, i}^l, \vec{q}_{n, i}^l, \vec{v}_{n, i}^l \in \mathbb{R}^{d_k}$ are called the key, query, and value vectors, respectively.

The core of self-attention lies in computing the \textit{attention weights}
\begin{equation}
    \alpha_{n n', i}^l = \text{softmax}\left( \frac{(\vec{q}_{n, i}^l)^T \cdot \vec{k}_{n', i}^l}{\sqrt{d_k}} + M_{n n', i}^l \right),
    \label{eq:attn_weight}
\end{equation}
where $M_{n n', i}^l$ is the masking matrix. The softmax function, defined as:
\begin{equation}
    \text{softmax}(\vec{z}) = \frac{1}{\sum_{j=1}^K e^{z_j}} (e^{z_1}, e^{z_2}, \dots, e^{z_K}),
    \label{eq:softmax}
\end{equation}
where $\vec{z} = (z_1, z_2, \dots, z_K)$, assigns a normalized probability to each entry of the input. Intuitively, Eq.~\ref{eq:attn_weight} assigns an ``attention score'' to each pair $(n, n')$ and captures how much influence the key at position $n'$ has on the query at position $n$. 

Crucially, since the Lanczos coefficients are causal, we implement a \textit{masked} self-attention mechanism through $M_{n n', i}^l$. Specifically, we set:
\begin{equation}
    M_{n n', i}^l = 
\begin{cases}
    0 \qquad  &n' \leq n \\
    -\infty  \qquad &n' > n,
    \label{eq:mask}
\end{cases}
\end{equation}
in all layers $l$ and attention heads $i$. Combined with the definition of softmax (Eq.~\ref{eq:softmax}), this mask ensures that the prediction at index $n$ depends \textit{only} on previous tokens ($n' \leq n$), so that the physical nature of the problem is respected. 

Finally, for each head, the attention weights are used to produce a weighted sum of the value vectors:
\begin{equation}
   \vec{x}_{n, i}^l = \sum_{n'} \alpha_{n n', i}^l \vec{v}_{n', i}^l.
\end{equation}

The output vectors $\vec{x}_{n,i}^l$ from each of the $H$ attention heads are then concatenated to form a single vector $\vec{x}_n^l \in \mathbb{R}^{d_\text{model}}$. The set of output vectors from self-attention, $\{\vec{x}_1^l, \vec{x}_2^l, \dots, \vec{x}_m^l \}$, is next processed by a position-wise Feed-Forward Network (FFN). This sub-layer is applied to each position $n$ independently and identically, and it allows the model to capture complex non-linear dependencies in the Lanczos sequence that the linear attention mechanism might miss. The FFN consists of two linear transformations with a non-linear activation function $\sigma(\cdot)$ in between:
\begin{equation} 
\text{FFN}(\mathbf{x}_n^l) = \sigma(\mathbf{x}_n^l W_1 + \mathbf{b}_1) W_2 + \mathbf{b}_2.
\end{equation}

In our proposed architecture, the hidden dimension of the FFN is chosen to be $4 \times d_{\text{model}}$. Both the self-attention and FFN sub-layers are encapsulated by residual connections and layer normalization, to ensure numerical stability during training.

Finally, the next-token prediction (namely, the next difference in Lanczos coefficients $\Delta b_{m+1}$), is produced by a learned linear ``unembedding'' (output) layer applied to the last hidden state $ \vec{h}_{m}^L$. Concretely, we predict
\begin{equation}
\Delta b_{m+1}
= W_O\mathbf{h}_{m}^{L} + b_O,
\quad
W_O \in \mathbb{R}^{1\times d_{\text{model}}},\ b_O\in\mathbb{R}.
\end{equation}

The learnable parameters of the transformer model consist of parameters in the embedding layer (weights and biases $W_E, \vec{b}_E$); in the decoder blocks ($W_K, W_Q, W_V, W_1, W_2, \vec{b}_1, \vec{b}_2$ for \textit{all} layers and \textit{all} attention heads); and in the un-embedding layer ($W_O, b_o$). Each transformer model $f$ is parameterized by $\Theta \in \mathbb{R}^P$, where $P$ is the total number of learnable parameters in the transformer model. We train the model by finding the set of model parameters $\Theta^*$ that minimizes the next-step prediction error on the increment sequence:
\begin{equation}
    L(\Theta)= \sum_{m=n_{\mathrm{in}}}^{T-1}
\left(\Delta b_{m+1}^{(j)}-f_\Theta\!\left(\Delta b_{1}^{(j)}, \Delta b_{2}^{(j)}, \dots \Delta b_{m}^{(j)}\right)\right)^2
\label{eq:loss}
\end{equation}
averaged over all Lanczos sequences (labeled by $j$) in the training set. Here $f_\Theta(\cdot)$ denotes the output of the transformer model. In Appendix~\ref{app:hyperparameters}, we detail the hyperparameters used in the transformer model, as well as ones used during training.

Once trained, the transformer model acts as a generative sequence predictor: at inference, it takes the prefix $\Delta b_1, \Delta b_2, \dots, \Delta b_{\nin}$ as the ``prompt'' and predicts $\Delta b_{\nin+1}$, which is then appended to the end of the input sequence. $\{\Delta b_1, \dots, \Delta b_{\nin+1} \}$ is then again input into the transformer to predict $\Delta b_{\nin+2}$. This process is repeated iteratively, thereby reconstructing long Lanczos sequences.

To benchmark the proposed transformer's performance and provide a baseline, we fit the first $\nin$ coefficients to asymptotic forms motivated by the UOGH, augmented by an even–odd staggering term. This form is the standard baseline fit and has been widely adopted in previous works~\cite{MBL, approx_Greens_extrap_and_cutoff_spin, dynamics_oneandtwo_d,pseudomode_extrapolate_spin, Yates_2020, Yates2_staggering, Yates_Floquet_staggering}. Concretely, we fit the first $\nin$ coefficients to:
\begin{align}
    b_n = \alpha n/\log(n) + \gamma  + \gamma^* (-1)^n \,\,\,\,\,\, d=1; \label{eq:baselinefit_d1} \\
    b_n = \alpha n + \gamma  + \gamma^* (-1)^n \,\,\,\,\,\, d \neq 1, \label{eq:baselinefit_dneq1}
\end{align}
where $d$ denotes the spatial dimension of the system. We obtain the optimal set of parameters $\alpha, \gamma, \gamma^*$ through least-squares regression. We extrapolate by evaluating the fitted expression on $n > \nin$.

As a quantitative measure of how much error the linear extrapolation and transformer predictions make, we evaluate the Root-Mean-Squared Error (RMSE) at each extrapolated index $n$, defined as:
\begin{align}
    \text{RMSE}(n) &= \sqrt{ \frac{1}{N_T} \sum_{i=1}^{N_T} \left( b_n^{(i)} - b_n^{\text{pred}; (i)}  \right)^2}
    \label{eq:RMSE}
\end{align}
where $N_T$ denotes the number of testing samples averaged over, and $b_n^\text{pred}$ is the Lanczos coefficient at index $n$, as predicted by either the transformer model or the linear fit.

\section{Numerical Results \label{sec:results}}
\subsection{Classical Example: XYZ Top}
In this section, we begin by demonstrating the transformer's applicability to extrapolating Lanczos coefficients in a classical chaotic Hamiltonian. The classical XYZ spin top is an ideal testbed for our approach in two ways. First, compared to quantum many-body systems, it is computationally inexpensive to obtain the Lanczos coefficients for a single classical spin. More importantly, Lanczos sequences in classical systems do not saturate because, unlike quantum systems, there is no constraint from finite Hilbert-space dimension~\cite{UOGH, MBL}. It is therefore straightforward to compute long sequences of coefficients to use for training, which allows for assessing the transformer's performance at prediction horizons ten times the input window.

The Hamiltonian we consider is the classical XYZ model, a paradigmatic system for studying classical chaos~\cite{LiuMuellerXYZ}:
\begin{equation}
    H_{\text{XYZ}} = J_x x^2 + J_y y^2 + J_z z^2,
    \label{eq:HXYZ}
\end{equation}
where $x, y, z$ denote the components of a classical spin.

As discussed in Sec.~\ref{sec:learning}, we fix the initial operator to $O_0 = z$. We sample each of $J_x, J_y, J_z$ independently from a uniform distribution on $[0,1]$. Each set of couplings specifies an instance of XYZ Hamiltonian, for which we compute the Lanczos coefficients up to $T=100$. Details for performing the computation are included in Appendix.~\ref{app:classical_lanczos}.

In Fig.~\ref{fig:classical}, we show the transformer's ability to forecast coefficients, using an input window of length  $\nin=10$. As a baseline, we also include
predictions from the linear asymptotic fit in Eq.~\ref{eq:baselinefit_dneq1}. In Fig.~\ref{fig:classical}(a), we show a representative sequence of the actual $b_n$'s for an XYZ Hamiltonian \textit{not} included within the transformer's training data, as well as the extrapolated coefficients by the transformer and linear fit, at an index horizon ten times the training window. The transformer remains consistently closer to the ground truth, and its advantage is even more pronounced deep into the extrapolation regime (see inset).

To quantify performance of the two approaches across instances, we also average the squared prediction errors of $100$ unseen instances of $H_{\text{XYZ}}$ for both methods. The resulting RMSE (Eq.~\ref{eq:RMSE}) is shown in Fig.~\ref{fig:classical}(b). The transformer achieves substantially lower error, often by a few times, compared to the linear extrapolation baseline.

\begin{figure}
    \centering
    \includegraphics[width=0.75\linewidth]{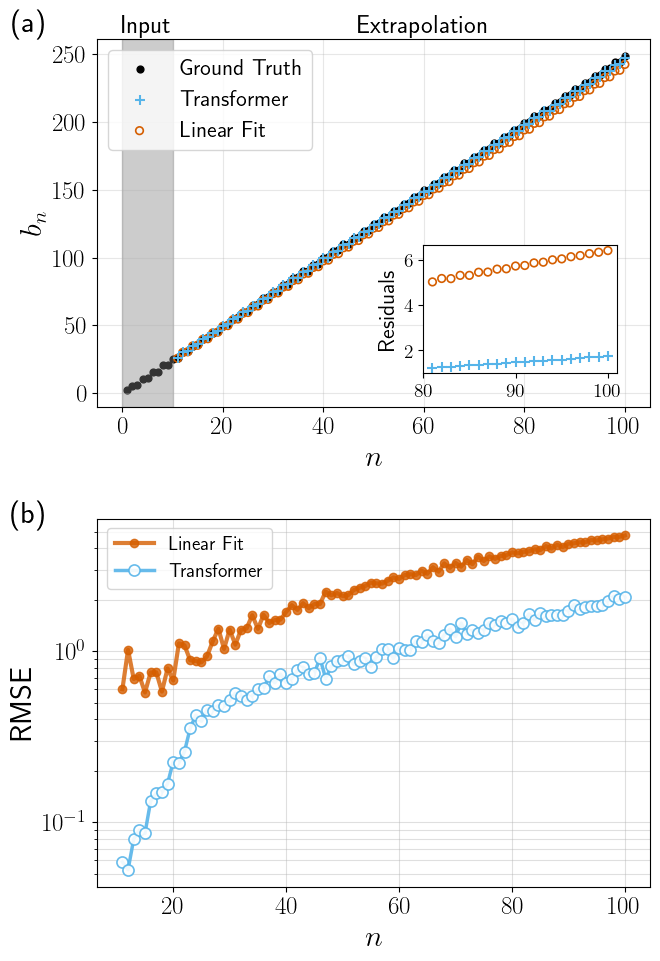}
    \caption{Transformer performance on the classical XYZ model. \textbf{(a)} Extrapolation of Lanczos coefficients from the input interval $n \leq 10$ to $n=100$. The transformer's predictions remain closer to the ground truth, whereas the linear-fit baseline exhibits a systematic bias as $n$ increases. \textit{Inset:} Residuals $b_n - b_n^{\text{pred}}$ deep in the extrapolation regime. The transformer's error remains a few times smaller than linear extrapolation. (b) Root Mean Squared Error (RMSE) averaged over 100 unseen instances of $H_{\text{XYZ}}$. The transformer achieves an error a few times smaller compared to the linear baseline, effectively capturing the subleading dynamics of the coefficients.}
    \label{fig:classical}
\end{figure}

\subsection{Quantum Example: Transverse-Field Ising Model \label{sec:TFIM}}
With the validity of our approach tested on the classical model, next we turn to quantum systems.
Compared to those in classical chaotic systems, the Lanczos coefficients under chaotic quantum Hamiltonians generally exhibit larger deviations from the asymptotic linear growth~\cite{UOGH, deviation1, deviation2}.
Furthermore, generating Lanczos sequences is significantly more challenging for larger system sizes, particularly due to the memory cost associated with the exponential growth of Hilbert space dimension.
In this section, we show that the transformer is not only capable of capturing subtle, long-range correlations between $b_n$'s, but also of forecasting coefficients from larger system sizes, \textit{without retraining}.

The quantum system we consider is the Transverse-Field Ising Model (TFIM) with longitudinal (integrability-breaking) field, an archetypal model for studying non-integrable dynamics and scrambling~\cite{TFIM, TFIM_lanczos}:
\begin{equation}
    H = \sum_i J Z_i Z_{i+1} + g X_i + h Z_i,
\end{equation}
where $i$ labels the site and $X_i$ and $Z_i$ denote the Pauli-$x$ and $z$ operators, respectively. A non-zero longitudinal field $h \neq 0$ breaks integrability and leads to chaotic operator growth, for which the one-dimensional UOGH motivates the asymptotic form (Eq.~\ref{eq:UOGH}).

To train the transformer, we fix the initial operator to $O_0 = \sigma_1^z$. We use open boundary conditions and set $J = 1$. We sample the dimensionless parameters $g/J$ and $h/J$ uniformly from $[1.0, 2.0]$
and $[0.1, 1.0]$, respectively. Because $h \neq 0$ breaks integrability, all sampled Hamiltonians are in the chaotic regime. Throughout training, the system size is fixed to be $L=8$ and the sequence length to be $T=30$. Throughout the main text, we use $\nin=10$ input coefficients for training. However, fewer coefficients may be used to achieve accurate forecasting; see Appendix~\ref{app:nin}.

\begin{figure}
    \centering
    \includegraphics[width=0.75\linewidth]{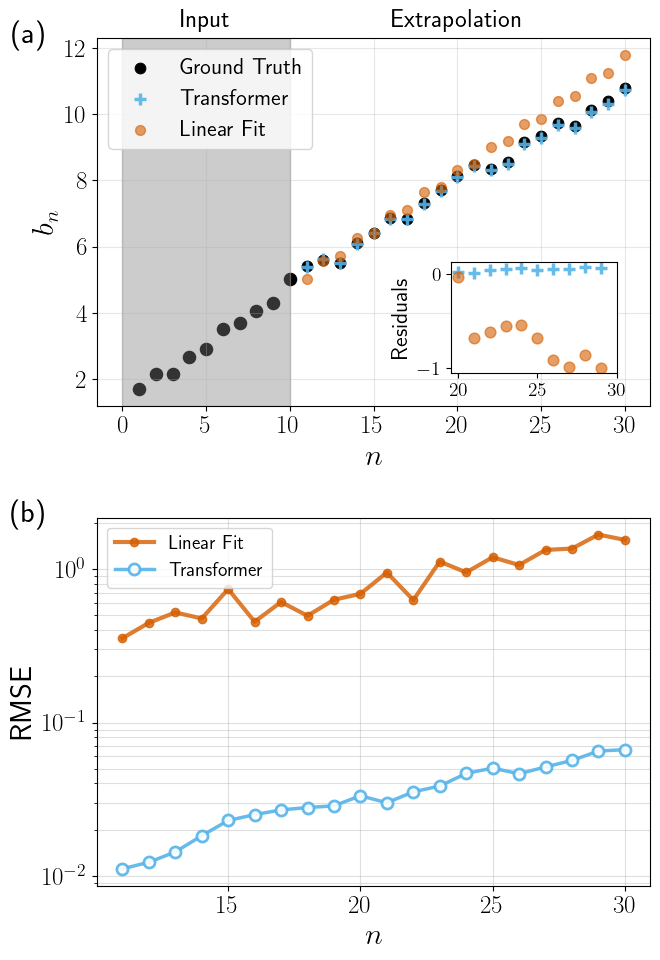}
    \caption{Transformer performance on the quantum Transverse-Field Ising Model (TFIM). \textbf{(a)} Extrapolation of Lanczos coefficients for an unseen $L=8$ system. \textit{Inset:} Residual $b_n-b_n^{\text{pred}}$ for the transformer and asymptotic fit at later $n$. \textbf{(b)} Root Mean Squared Error (RMSE) averaged over $100$ unseen test sequences, demonstrating that the transformer systematically outperforms the asymptotic fit by capturing subleading, non-linear deviations from the asymptotic ramp.}
    \label{fig:TFIM_L8}
\end{figure}

We first evaluate performance on unseen Hamiltonians drawn from the same parameter distribution. Fig.~\ref{fig:TFIM_L8}(a) shows a representative instance: the transformer extrapolation tracks the exact Lanczos coefficients closely well beyond the input window, while the $d=1$ baseline fit (Eq.~\ref{eq:baselinefit_d1}) develops a systematic deviation as $n$ increases. In Fig.~\ref{fig:TFIM_L8}(b), we further quantify this trend by the root-mean-squared error (RMSE) averaged over an ensemble of 100 unseen Lanczos sequences. Across the full extrapolation regime, the transformer typically achieves an order-of-magnitude reduction in error relative to the baseline fit, indicating that it learns history-dependent, subleading structures in $\{b_n\}$ that are not captured by asymptotic fits.

Having established accurate extrapolations of $\{b_n\}$ from short input windows, we next assess the transformer's utility in reconstructing physically relevant quantities. We focus on the Krylov complexity $K(t)$ and autocorrelation function $C(t)$, defined in Eqs.~\ref{eq:krylov_complexity} and \ref{eq:autocorrelation} respectively. Specifically, for each test instance, we consider three sequences of length $T$: the ground truth coefficients; and the sequences extrapolated by the transformer and asymptotic fit from the same initial prefix of $\nin$ coefficients. For each sequence, we solve the tight-binding problem (Eq.~\ref{eq:hoppingeq}) on the Krylov chain by exponentiating the Liouvillian (Eq.~\ref{eq:L_tridiag}) to obtain $K(t)$ and $C(t)$, as well as their counterparts $K^{\text{pred}}(t)$ and $C^{\text{pred}}(t)$ using extrapolated coefficient sequences. We consider the test-set RMSE of the reconstructed observables. Concretely, we define
\begin{align}
    |\Delta K(t)| &= \sqrt{ \frac{1}{N_T} \sum_{i=1}^{N_T} |K_i(t) - K_i^{\text{pred}}(t)|^2} \, \,,
    \label{eq:deltaKt}
\end{align}
where $N_T$ is the number of testing sequences, and $|\Delta C(t)|$ is defined analogously.

\begin{figure}
    \centering
    \includegraphics[width=1\linewidth]{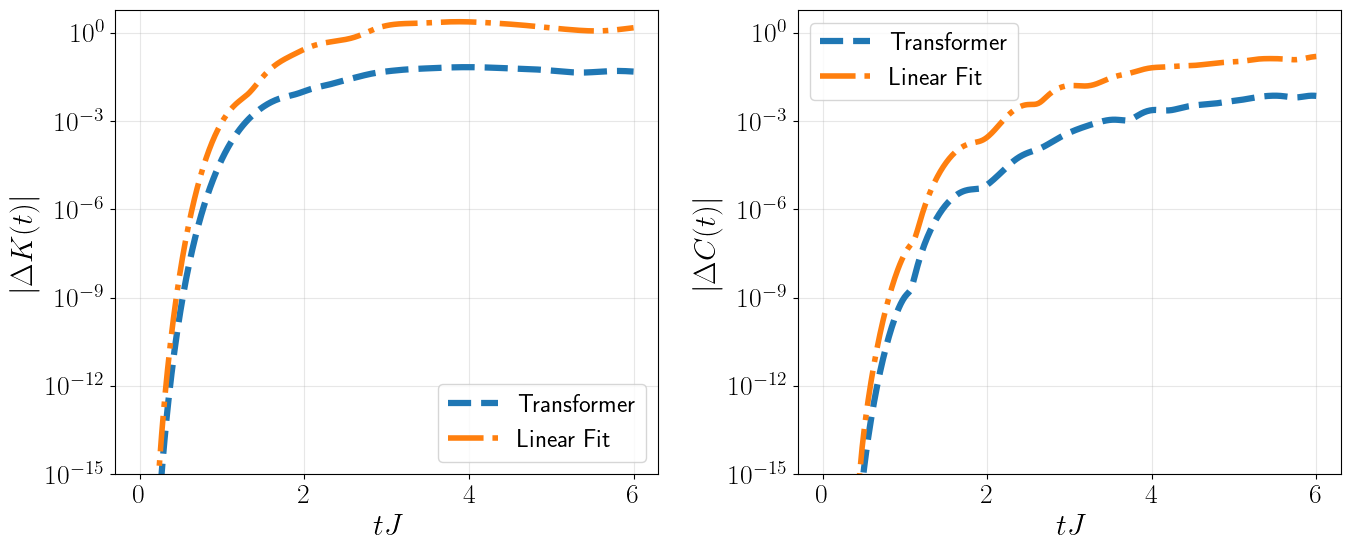}
    \caption{Root-mean-squared error (Eq.~\ref{eq:deltaKt}) in the reconstruction of Krylov complexity and autocorrelation function, using extrapolated Lanczos sequences by transformer and asymptotic fit, compared to reconstructions from the exact coefficients. Errors are averaged over $N_T = 100$ test sequences. While both methods are comparably accurate at short times $tJ \lesssim 1$, transformer-based extrapolation leads to substantially smaller errors at later times by orders of magnitude.}
    \label{fig:observables}
\end{figure}

In Fig.~\ref{fig:observables}, we compare the errors of both methods. At short times ($tJ \lesssim 1$), both methods reconstruct $K(t)$ and $C(t)$ with  similar accuracy. However, at later times (roughly $tJ \gtrsim 2$), the reconstructed quantities from the transformer consistently achieve errors orders of magnitude smaller than the respective quantities obtained from coefficients by  asymptotic fits. The model's superior performance at observable level therefore provides evidence that the transformer's fidelity in coefficient forecasting translates directly into more accurate predictions of physical dynamics. Our observation also reinforces the motivation that small but systematic errors in the extrapolated tail of $\{b_n\}$ could be responsible for such deviations in the Krylov-chain dynamics.

\begin{figure}
    \centering
    \includegraphics[width=0.7\linewidth]{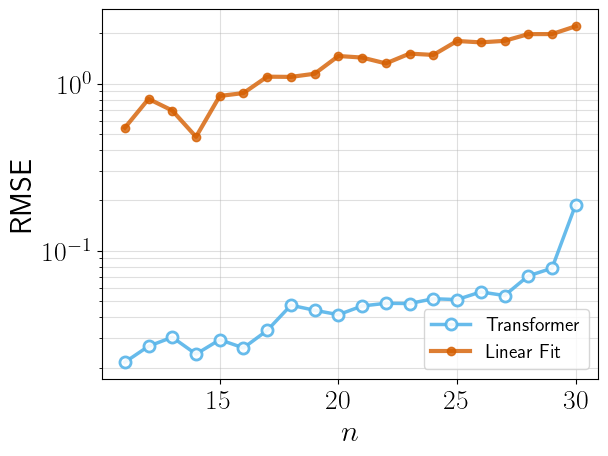}
    \caption{Zero-shot transferability of the transformer model. Despite being trained on data from an $L=8$ chain of transverse-field Ising model, the transformer is able to accurately forecast Lanczos sequences from a larger system size $L=12$. The RMSE, averaged over $N_T=100$ sequences, remains consistently smaller than those from baseline fits.}
    \label{fig:L12}
\end{figure}

Remarkably, the transformer also generalizes \textit{beyond} the conditions represented in its training data. Fig.~\ref{fig:L12} shows that a model trained exclusively on $L=8$ Lanczos sequences can extrapolate coefficients for larger systems ($L=12$) from short input windows, \textit{without retraining}. While the transformer predictions for $L=12$ are generally less accurate than the in distribution, $L=8$ case (Fig.~\ref{fig:TFIM_L8}(b)), they still consistently outperform the asymptotic fit by roughly an order of magnitude.

This transferability across system sizes of the transformer model substantially strengthens the practical value of our approach. In particular, one can collect training data at smaller sizes, where Lanczos computations are feasible and considerably cheaper, to train a single transformer model. That same model can then be deployed to larger system sizes, for which direct iterations can become prohibitively expensive.

\subsection{Integrable Example: Quantum XXZ Model}
The systems considered in the previous sections are chaotic, for which the UOGH (Eq.~\ref{eq:UOGH}) is expected to apply. In contrast, integrable systems have qualitatively different scaling of Lanczos coefficients~\cite{UOGH}: specifically, the Lanczos coefficients $b_n$ typically exhibit \textit{sub-linear} growth. Since the asymptotic linear fit $b_n \sim \alpha n + \gamma$ is no longer valid, one natural question is whether the transformer model can still extrapolate the Lanczos coefficients accurately.

To test this, we consider the spin-half Heisenberg XXZ model, with the Hamiltonian:
\begin{equation}
    H_\text{XXZ} = \sum_i JX_i X_{i+1} + J Y_i Y_{i+1} + \Delta Z_i Z_{i+1},
\end{equation}
which is integrable and exactly solvable via the Bethe ansatz~\cite{XXZ}.

Similar to the non-integrable TFIM in Sec.~\ref{sec:TFIM}, we fix $J=1$ and sample $\Delta$ from a uniform distribution on $[0.5, 2.0]$. The initial seed operator is $O_0 = Z_1$ (the Pauli-$Z$ operator on the first site). We impose open boundary conditions on a one-dimensional chain of length $L=8$. The number of training samples, as well as parameters of the trained transformer model, are all identical to the quantum TFIM case, as listed in Appendix~\ref{app:hyperparameters}.

As shown in Fig.~\ref{fig:integrable}(a), the Lanczos coefficients grow sub-linearly. Despite the non-linear trend, the trained transformer model captures the growth accurately, with minimal error even deep into the extrapolation regime. We note that 
integrable models generally exhibit more pronounced even-odd staggering in the Lanczos coefficients compared to chaotic systems~\cite{UOGH, subleading3}.

Figure~\ref{fig:integrable}(b) shows the RMSE (Eq.~\ref{eq:RMSE}) averaged over $N_T = 100$ unseen test sequences. The error remains below $10^{-2}$ across the full extrapolation range without systematic growth. These results demonstrate that the transformer captures sequence structure well beyond monotone linear growth. In particular, the even-odd staggering is reproduced without being built into the model as an explicit fitting parameter.

We highlight that unlike the chaotic case, there is no well-established closed-form asymptotic fit to serve as a baseline for integrable models. Indeed, how $b_n$ depends on $n$ in integrable systems can be non-universal and model-dependent~\cite{UOGH}. The transformer, by contrast, is agnostic to the functional form of the asymptote and learns the sequence structure directly from data.

\begin{figure}
    \centering
    \includegraphics[width=0.75\linewidth]{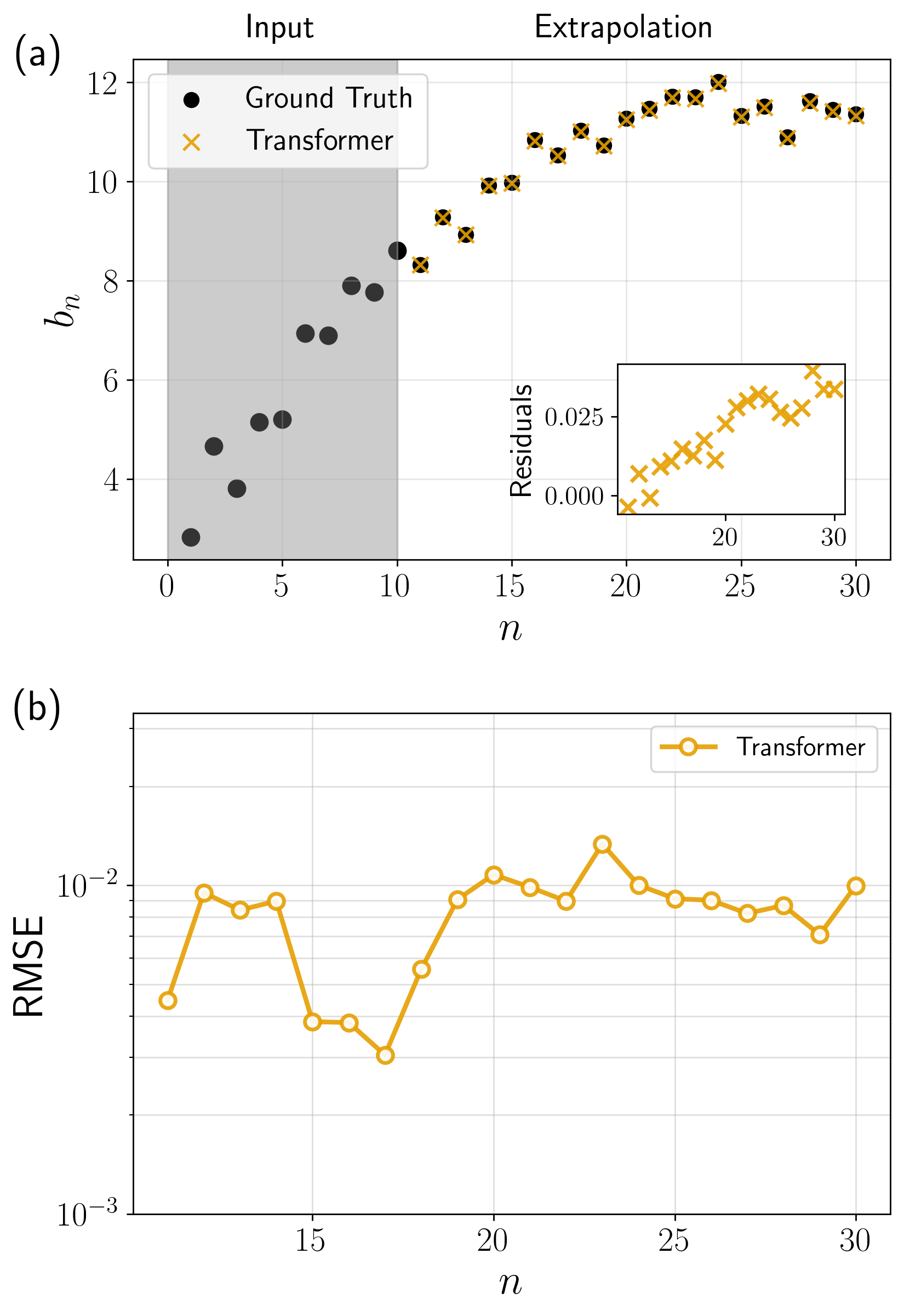}
    \caption{Transformer performance on the integrable Heisenberg XXZ model.
    \textbf{(a)} An example of extrapolated sequence of Lanczos coefficients from the input prefix $n \leq 10$ (shaded) to $n = 30$. The ground-truth coefficients exhibit sub-linear growth with pronounced even-odd staggering, both of which the transformer reproduces accurately.
    \textit{Inset:}~Residuals $b_n - b_n^\text{pred}$ in the extrapolation regime, remaining below $0.03$ throughout the extrapolation window.
    \textbf{(b)}~Root-mean-squared error (Eq.~\ref{eq:RMSE}) averaged over $N_T = 100$ unseen test sequences, showing errors below order $10^{-2}$, without systematic growth across the extrapolation window.}
    \label{fig:integrable}
\end{figure}

\section{Attention Map and Ablation Studies \label{sec:discussion}}

One of the major advantages of the transformer architecture, compared to other neural-network models, lies in its \emph{interpretability}. In particular, the attention weight matrix provides a mechanistic view of which parts of the input sequence are emphasized by the trained transformer model.

\begin{figure}
    \centering
    \includegraphics[width=1\linewidth]{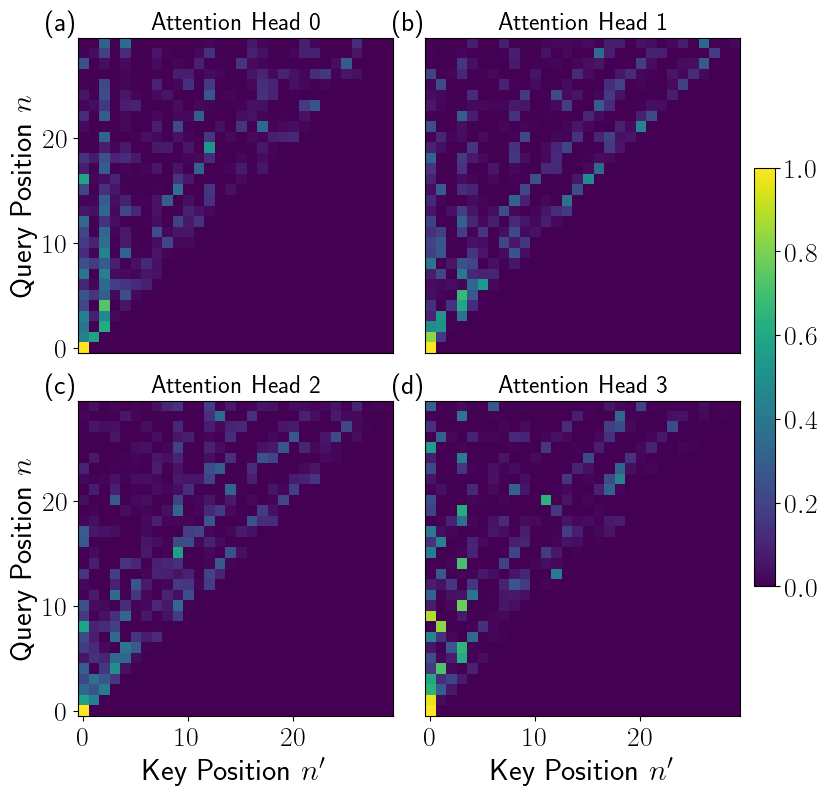}
    \caption{Batch- and layer-averaged attention maps for the four attention heads. The triangular support confirms causal masking. Across heads, attention concentrates near the diagonal, exhibits a checkerboard pattern, and assigns significant weight to the earliest coefficients. These patterns reveal that the model’s prediction relies on both early entries and recent coefficients, while effectively capturing the even-odd staggering of the Lanczos coefficients.}
    \label{fig:attention_map}
\end{figure}

In this section, we first analyze the self-attention weights $\alpha_{n n', i}^l$ in the model trained on Lanczos sequences from the transverse-field Ising model. We average over all decoder layers $l$ to ensure the result is representative of the whole network. Fig.~\ref{fig:attention_map} shows the attention map for each head, also averaged over a batch of 64 sequences. In our notation, $\alpha_{n n', i}$ (Eq.~\ref{eq:attn_weight}) measures the attention weight (in attention head $i$) from the \emph{key} at position $n'$ to the \emph{query} at
position $n$ i.e., quantifying how much the model attends to the coefficient at step $n'$, when predicting its value at step $n$). Because we employ causal masking, $\alpha_{n n'}=0$ for $n' \ge n$ after the softmax operation. The observed triangular structure in all four heads provides a direct check that the causal constraint is indeed being enforced in the trained model.

Beyond this structure, we observe three robust patterns of attention across all heads. First, the attention concentrates along the main diagonal $n' \lesssim n$, indicating that the model relies strongly on the most recent coefficients when forming its next-step prediction. In addition, the attention maps consistently display a staggered (checkerboard) pattern, indicating sensitivity to the even-odd structure of the Lanczos sequence. Note that unlike the linear fits, Eqs.~\ref{eq:baselinefit_d1} and~\ref{eq:baselinefit_dneq1}, the even-odd staggering is not built-in, but in fact learned by the transformer. Finally, a wide range of query positions assign significant weight to the first few coefficients across heads, visible as bright vertical lines at small key positions. This suggests that early coefficients serve as a kind of ``anchor'' that subsequently enables global property inference of the sequence, even for predictions far beyond the input window.

The observed attention patterns naturally motivate an inference-time \textit{ablation study}: investigation into the model's performance when we suppress selected parts of the attention pattern while keeping all weights fixed. Such a study provides insight into how each part of the attention mechanism contributes to the model's predictions.
Concretely, we compare the performance of the full model against three modified attention masks at inference. We set the mask (Eq.~\ref{eq:mask}) $M_{nn', i} = -\infty$ (and therefore $\alpha_{nn', i} = 0$, after the softmax operation) when: $n$ and $n'$ have different parity (``parity'' masking); $n' \le n-k$, so that each query position $n$ may attend only to the most recent
$k$ preceding coefficients, thereby removing long-range attention (``long range'' masking); and $n' \le k$, where we remove access to the first $k$ coefficients (``early'' masking). We restrict to $k=3$ in this work.

\begin{figure}[t]
    \centering
    \includegraphics[width=0.7\linewidth]{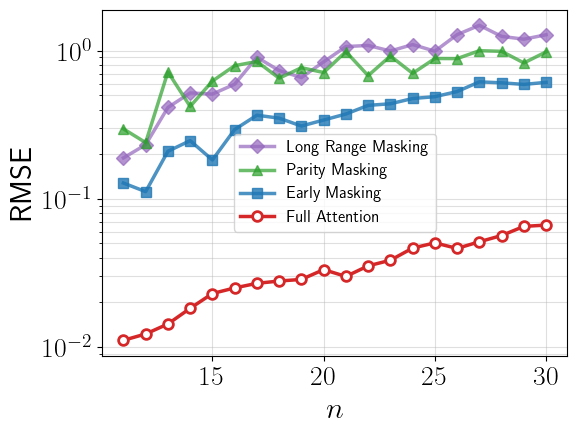}
    \caption{Ablation study on the attention weight matrix. Comparison of the Root Mean Squared Error (RMSE), averaged over 100 sequences, for Lanczos coefficient extrapolation using the full transformer model versus models with restricted attention masks. The model with full attention outperforms its counterparts whose attention is restricted to local windows, same-parity history, or where the first three coefficients are hidden. This confirms that long-range dependencies and early-time coefficients are essential for accurate forecasting.}
    \label{fig:ablation}
\end{figure}

In Fig.~\ref{fig:ablation}, we compare the model's performances in all four cases: full attention against the three ways of masking. The resulting RMSEs, averaged over $100$ unseen Lanczos sequences, demonstrate that all three ablations downgrade the prediction accuracy significantly. In particular, restricting the model
to only the past $k=3$ coefficients produces an order-of-magnitude increase in RMSE across the extrapolation regime, indicating that long-range attention is indeed essential for accurate forecasting. Even-odd masking and earliest token removal also lead to significantly large errors, consistent with the visual checkerboard and vertical striped patterns in Fig.~\ref{fig:attention_map}, indicating that even-odd staggering and early coefficients provide a global reference for coefficient predictions.

The ablation study demonstrates that attention across the entire window of inputs is crucial for model performance. From a physics perspective, the importance of long-range context is expected. The Lanczos recursion (Eq.~\ref{eq:lanczos}) constructs new elements $|O_n)$ of the Krylov basis from $|O_{n-1})$, $|O_{n-2})$, as well as the preceding Lanczos coefficient $b_{n-1}$. The iterative nature of this procedure means that each subsequent coefficient $b_n$ implicitly aggregates information propagated from the \textit{entire} prior history of the recursion. These observations directly support the central motivation of our work: that Lanczos coefficients form a
causal sequence with long-range, structured correlations, and model architectures that leverage attention across the full input window can capture subleading structure that is missed by purely asymptotic fits.

\section{Summary and Outlook\label{sec:conclusion}}
In this work, we introduce a deep-learning approach to probing quantum dynamics through Krylov space. We treat the Lanczos coefficients as a causally-related sequence, and use the transformer architecture to learn the intricate, non-linear correlations between coefficients. The results show that our data-driven method significantly outperforms standard asymptotic fits, achieving orders-of-magnitude improvements in forecasting accuracy of both the Lanczos coefficients and physical observables. Remarkably, the model can be trained on data from smaller system sizes and be transferred to extrapolate sequences from a larger system. By probing the inner workings of the attention mechanism, we discover that the model learns subleading structures in the sequences to make accurate predictions.

Despite its accuracy for coefficient extrapolation, the present model is limited in two ways. First, the transformer is trained on Lanczos sequences generated from a given family of Hamiltonians and a fixed initial operator ($\mathcal{O}_0 = Z_1$ for the quantum models). Since Lanczos coefficients depend on the seed operator and Hamiltonian, a model trained on one family and one initial operator cannot, in general, transfer to sequences starting from a different physical model or initial seed operator. Extending the model to generalize across models and operators is an interesting direction for future work. Second, the training data are
drawn from a region of parameter space corresponding to a single regime (either chaotic or integrable). Generalization across phase boundaries would most likely
require a model conditioned on the Hamiltonian parameters (e.g. feeding the network Hamiltonian parameters along with the coefficient prefix.)

Given these limitations, there are several natural directions for future work. 
One interesting direction is to build towards a ``foundation model'' for quantum dynamics~\cite{NOQS, UNP}, but formulated in Krylov space. While we have demonstrated the approach's capability to transfer across system sizes within a given family of Hamiltonians, a more ambitious goal would be to train a single model that generalizes across different Hamiltonian families and initial operators. Such a model could potentially serve as a general-purpose surrogate for Lanczos coefficient extrapolation across a broad class of quantum many-body systems and initial conditions. Whereas Refs.~\cite{NOQS, UNP} build foundation models of quantum dynamics from the perspective of state evolution, such a Krylov-space model could serve as a complementary approach in the Heisenberg (operator-evolution) picture.

Moreover, our approach could potentially serve as a bridge between experiments and the theory of operator growth.
As discussed in Sec.~\ref{sec:lanczos}, the Lanczos coefficients are directly related to the autocorrelation function $C(t)$. Therefore, measuring $C(t)$ provides a potential avenue for extracting the first few coefficients experimentally.
Specifically, one can compute the spectral \textit{moments} $\mu_{2n}$
\begin{equation}
    \mu_{2n} = (O_0 | \mathcal{L}^{2n} | O_0) = (-1)^n \frac{d^{2n}}{dt^{2n}}C(t)|_{t=0}
\end{equation}
from $C(t)$. The Lanczos coefficients $b_n$ can be efficiently computed from a recursion algorithm performed on the spectral moments (see Appendix A of Ref.~\cite{UOGH}). Conversely, given the set of coefficients $\{ b_n\}$, one can reconstruct $C(t)$ by solving the tight-binding problem on the Krylov chain (Eq.~\ref{eq:hoppingeq}).

In practice, experimental noise and limited resolution may limit the number of moments (and therefore the number of Lanczos coefficients) that can be reliably computed. Our machine-learning based model, then, provides a natural tool for extending the horizon in Krylov space.

\section*{Acknowledgments}
We acknowledge Junkai Dong, Yang Peng, Daniel Parker, and Hsiu-Chung Yeh for helpful comments. We acknowledge the use of large language models (ChatGPT 5.2 and Sonnet 4.6) for research assistance. ZQ thanks Xiangyu Cao and Thomas Scaffidi for collaboration on a previous project. ZQ gratefully acknowledges support from the Simons Center for Geometry and Physics, Stony Brook University at which some of the research for this paper was performed during the workshop \emph{Complexity, Information, and Tractable Simulations of Quantum Many-body Dynamics}.

\bibliography{main}

\appendix
\section{Lanczos Algorithm for Classical Chaotic Systems}
\label{app:classical_lanczos}
In Sec.~\ref{sec:lanczos}, we review the Lanczos algorithm for quantum operators acting on finite-dimensional Hilbert spaces. The transition to dynamics in a classical phase space involves replacing the commutator with the Poisson bracket and the infinite-temperature trace with a phase-space average. In the classical limit, observables are no longer finite dimensional operators, but functions of the phase space variables. 

For a single classical spin, the dynamics occur on the surface of a unit sphere $S^2$. The phase space is naturally spanned by the spherical harmonics $Y_l^m(\theta, \phi)$: any observable $A(x, y, z)$ can be expanded in that basis. The infinite-temperature trace is replaced by a normalized integral over the surface of the sphere:
    \begin{equation}
        (A|B) = \frac{1}{4\pi} \int_{0}^{2\pi} \int_{0}^{\pi} A^*(\theta, \phi) B(\theta, \phi) \sin\theta d\theta d\phi,
    \end{equation}
which ensures $(1|1) = 1$, the classical analogue of the quantum normalization $\text{Tr}(\mathbb{I})/\text{Tr}(\mathbb{I}) = 1$.

In the case of the XYZ top (Eq.~\ref{eq:HXYZ}), the variables are the components of the angular momentum vector $(x, y, z)$. To calculate the Lanczos coefficients, we first define the action of the Liouvillian super-operator $\mathcal{L}$. In the classical regime, the Heisenberg equation of motion becomes:
\begin{equation}
    \mathcal{L}|O) = \{H, O\}
\end{equation}
where $\{\cdot, \cdot\}$ denotes the Poisson bracket.

Following the procedure outlined in Sec.~\ref{sec:lanczos}, we compute the coefficients iteratively. Recall the Hamiltonian is $H=J_x x^2 + J_y y^2 + J_z z^2$ (Eq.~\ref{eq:HXYZ}). We start with the normalized operator $|O_0) \propto z$, such that $(O_0|O_0) = 1$. We calculate $|A_1) = \mathcal{L}|O_0) = \{H, O_0\}$. Using the chain rule $\{f^2, g\} = 2f\{f, g\}$:
    \begin{align}
        |A_1) &= J_x \{x^2, z\} + J_y \{y^2, z\} \nonumber \\
        &= 2J_x x \{x, z\} + 2J_y y \{y, z\} \nonumber \\
        &= 2J_x x(-y) + 2J_y y(x) = 2(J_y - J_x)xy
    \end{align}

In the last step, we used the fundamental brackets for the spin components $\{x, y\} = z$, $\{y, z\} = x$, and $\{z, x\} = y$.

The first coefficient is the norm $b_1 = \sqrt{(A_1|A_1)}$. Subsequent elements of the Krylov basis and Lanczos coefficients are calculated similarly, using the relation $|A_n) = \mathcal{L}|O_{n-1}) - b_{n-1}|O_{n-2})$ as well as the chain rule of Poisson brackets.

\section{Hyperparameters for Transformer Model \label{app:hyperparameters}}
The Transformer model used for extrapolating the Lanczos coefficients is a decoder-only architecture, as illustrated in Fig.~\ref{fig:transformer}. To ensure reproducibility, we provide the hyperparameter configuration used for Sec.~\ref{sec:results} in Table.~\ref{tab:hyperparameters}.

\begin{figure}[h!]
    \centering
    \includegraphics[width=\linewidth]{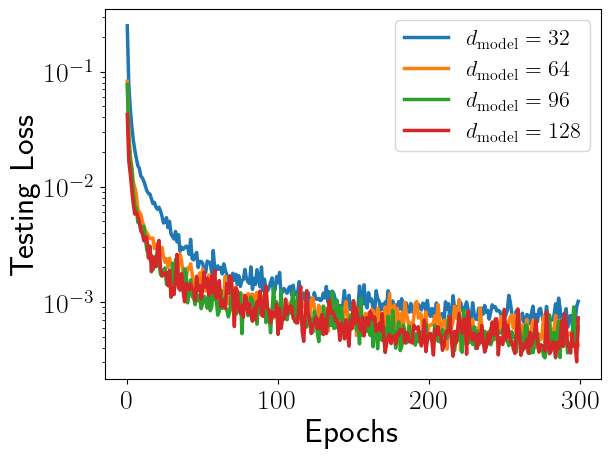}
    \caption{Testing loss during training, for various choices of $d_\text{model}$, with all other hyperparameters fixed. For $\d_\text{model} \geq 64$, the loss curves are indistinguishable and saturate with respect to the model dimension. On the other hand, a model with $d_\text{model}=32$ exhibits systematically higher test loss, indicating under-parameterization.}
    \label{fig:dmodel}
\end{figure}

For both classical and quantum systems, we choose an architecture with $L=3$ decoder layers and $4$ attention heads. This choice ensures that the network is sufficiently representative, while remaining simple and interpretable.
Indeed, transformer-based models with a few layers and attention heads have been sufficient in wavefunction representation~\cite{transformer_quantum_state, transformer_wavefunction, VMC_transformer}.
We fix the model dimension to be $d_\text{model} = 64$, which was chosen empirically. In Fig.~\ref{fig:dmodel}, we show the testing loss during training, for various choices of model dimension. Clearly, the loss curves sit on top of each other for $d_\text{model} \geq 64$, indicating saturation with respect to $d_{\text{model}}$, while the loss is visibly larger for $d_{\text{model}} = 32$, suggesting under-parameterization. The experiment demonstrates that a model dimension of 64 is sufficient to capture the essential features, while not introducing unnecessary parameters.

\begin{table}[b!]
\centering
\begin{tabular}{ll}
\hline
\textbf{} & \textbf{Hyperparameter} \\ \hline
\textbf{Architecture} & \\
Number of Decoder Blocks ($L$) & 3 \\
Embedding Dimension ($d_{\text{model}}$) & 64 \\
Number of Attention Heads & 4 \\
Feed-forward Network Dimension & $4 \times d_{\text{model}} = 256$ \\
Activation Function $\sigma$ & ReLU \\
Dropout Rate & 0.1 \\ \hline
\textbf{Data} & \\
Input Window Size ($\nin$) & 10 \\
Sequence Length & 30 \\
Positional Encoding & Sinusoidal~\cite{Vaswani2017} \\ 
Number of Training Samples & $10,000$ \\

\hline
\textbf{Training} & \\
Optimizer & AdamW~\cite{adamw} \\
Learning Rate & $1 \times 10^{-3}$ \\
Batch Size & 64 \\
Training Epochs & 300 \\ \hline
\end{tabular}
\caption{Hyperparameters for the Transformer-based extrapolation model. The parameters are the same across all three systems considered in Sec.~\ref{sec:results}.}
\label{tab:hyperparameters}
\end{table}

\section{Effect of Input Sequence Length $\nin$ \label{app:nin}}
Throughout Sec.~\ref{sec:results}, we have demonstrated the Transformer's ability to extrapolate Lanczos sequences using the first $\nin=10$ coefficients as inputs. However, due to the exponential growth of Hilbert space dimension, in practice, it can be challenging to collect even the first $\nin=10$ coefficients. Understanding how the performance depends on $\nin$ therefore affords an insight into the ``resource'' the network requires for accurate extrapolation.

Here we examine how the transformer's performance depends on $\nin$ by training separate models using $n_\text{in} \in \{6, 8, 10, 12\}$, keeping all other hyperparameters fixed. The training data and model architecture are identical to the TFIM case described in Sec.~\ref{sec:TFIM}: specifically, we aim to extrapolate to $n_\text{max}=30$, starting from inputs of varying lengths.
 
\begin{figure}[t!]
    \centering
    \includegraphics[width=1\linewidth]{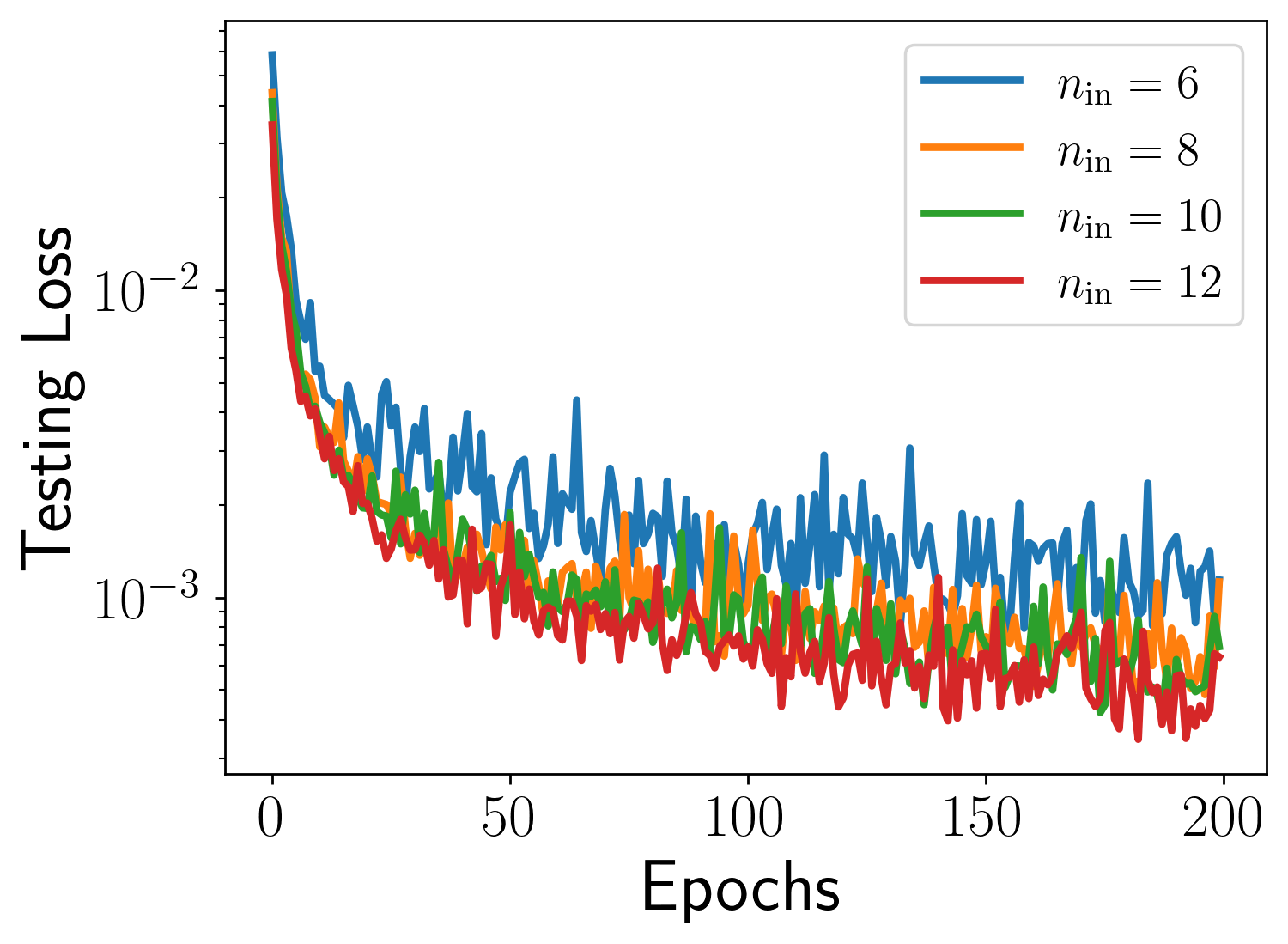}
    \caption{Testing loss during training for different input window lengths $n_\text{in} \in \{6, 8, 10, 12\}$, with all other hyperparameters fixed. For $n_\text{in} \geq 8$, the loss curves converge; even for $\nin=6$, the error remains on the same order of magnitude. This result indicates that a handful of input coefficients are sufficient for accurate extrapolation.}
    \label{fig:nin}
\end{figure}

Fig.~\ref{fig:nin} shows the testing loss during training for each choice of $n_\text{in}$. As expected, the loss is lower with more input coefficients. For $n_\text{in} \geq 8$, the loss curves are nearly indistinguishable, indicating that the transformer extracts sufficient information from as few as $8$ input coefficients to achieve accurate extrapolation. At $n_\text{in} = 6$, the loss is higher but remains within the same order of magnitude ($\sim 10^{-3}$), suggesting that even very short prefixes carry meaningful and sufficient signal.
 
These results indicate that the transformer's performance is robust to the choice of input window, and that $n_\text{in} = 10$ used in the main text is already in the saturated regime. This also demonstrates the practicality of our approach: in quantum many-body systems, where computing each additional coefficient becomes increasingly more expensive, a model can be trained to make accurate predictions with a handful of input coefficients.

\end{document}